\documentclass[lettersize,journal]{IEEEtran}
\usepackage{amsmath,amsfonts}
\usepackage{algorithmic}
\usepackage{array}
\usepackage[caption=false,font=normalsize,labelfont=sf,textfont=sf]{subfig}
\usepackage{textcomp}
\usepackage{stfloats}
\usepackage{url}
\usepackage{verbatim}
\usepackage{graphicx}
\usepackage{xcolor}

\def\BibTeX{{\rm B\kern-.05em{\sc i\kern-.025em b}\kern-.08em
    T\kern-.1667em\lower.7ex\hbox{E}\kern-.125emX}}
\usepackage{balance}
\begin{document}
\title{Noise Optimization for MKIDs with Different Design Geometries and Material Selections}

\author{%
Z. Pan, K. R. Dibert, J. Zhang, P. S. Barry, A. J. Anderson, A. N. Bender, B. A. Benson, T. Cecil, C. L. Chang, R. Gualtieri, J. Li, M. Lisovenko, V. Novosad, M. Rouble, G. Wang, and V. Yefremenko
\thanks{
(\textit{Corresponding author: Zhaodi Pan)}}
\thanks{Z. Pan is with the Argonne National Laboratory, 9700 South Cass Avenue, Lemont, IL, 60439, USA (email: panz@anl.gov)}%
\thanks{K. Dibert is with the University of Chicago, 5640 South Ellis Avenue, IL, 60637, USA (email: krdibert@uchicago.edu) }%
\thanks{J. Zhang, A. N. Bender, T. Cecil,  R. Gualtieri, J. Li, M. Lisovenko, V. Novosad, G. Wang, and V. Yefremenko are with the Argonne National Laboratory, 9700 South Cass Avenue, Lemont, IL, 60439, USA (emails: jianjie-zhang@outlook.com; abender@anl.gov; cecil@anl.gov; rgualtieri@anl.gov; juliang.li@anl.gov; mlisovenko@anl.gov; novosad@anl.gov; gwang@anl.gov; yefremenko@anl.gov)}%
\thanks{P. Barry is with Cardiff University, Cardiff CF10 3AT, UK (email: barryp2@cardiff.ac.uk)}
\thanks{A. J. Anderson and B. Benson are with Fermi National Accelerator Laboratory, PO BOX 500, Batavia, IL 60510, the University of Chicago, Chicago, IL, 60637, USA, and Kavli Institute for Cosmological Physics, U. Chicago, 5640 South Ellis Avenue, Chicago, IL, 60637, USA (email: adama@fnal.gov; kkarkare@kicp.uchicago.edu; bbenson@astro.uchicago.edu)}
\thanks{C. L. Chang is with the Argonne National Laboratory, Argonne, IL 60439 USA, the University of Chicago, 5640 South Ellis Avenue, Chicago, IL, 60637, USA, and Kavli Institute for Cosmological Physics, U. Chicago, 5640 South Ellis Avenue, Chicago, IL, 60637, USA (email: clchang@kicp.uchicago.edu) }%
\thanks{ M. Rouble is with McGill University, 845 Rue Sherbrooke O, Montreal, QC H3A 0G4, Canada (email: maclean.rouble@mcgillcosmology.ca)}
}

\markboth{Journal of \LaTeX\ Class Files,~Vol.~0, No.~0, April~2023}%
{Shell \MakeLowercase{\textit{et al.}}: Bare Demo of IEEEtran.cls for IEEE Journals}

\maketitle

\begin{abstract}
The separation and optimization of noise components is critical to microwave-kinetic inductance detector (MKID) development. We analyze the effect of several changes to the lumped-element inductor and interdigitated capacitor geometry on the noise performance of a series of MKIDs intended for millimeter-wavelength experiments. We extract the contributions from two-level system noise in the dielectric layer, the generation-recombination noise intrinsic to the superconducting thin-film, and system white noise from each detector noise power spectrum and characterize how these noise components depend on detector geometry, material, and measurement conditions such as driving power and temperature. We observe a reduction in the amplitude of two-level system noise with both an elevated sample temperature and an increased gap between the fingers within the interdigitated capacitors for both aluminum and niobium detectors. We also verify the expected reduction of the generation-recombination noise and associated quasiparticle lifetime with reduced inductor volume. This study also iterates over different materials, including aluminum, niobium, and aluminum manganese, and compares the results with an underlying physical model. 
\end{abstract}


\section{Introduction}
Future mm-wave experiments require ever-increasing detector counts to achieve the necessary sensitivity to pursue ambitious science cases.
Microwave kinetic inductance detectors (MKIDs) are well suited to large-format detector arrays, including photometer arrays \cite{calvo2016nika2, brien2018muscat, wilson2020toltec} and on-chip filter-bank spectrometers \cite{shirokoff2012mkid, endo2012development, barry2021design, mirzaei2020mu}. To achieve photon-limited sensitivity for each detector, it is necessary to understand each component of detector noise, including two-level system (TLS) noise, shot noise from the generation and recombination (GR) of quasiparticles, and amplifier noise. 

TLS noise is caused by the coupling of a resonator to a thin amorphous solid dielectric layer, where two-level tunneling states are thought to exist based on a phenomenological model \cite{phillips1972tunneling}. 
When atoms tunnel between two states, the resulting dipole can couple to the electric field of the resonator, causing excessive noise with a characteristic spectral shape proportional to $f^{-1/2}$, where $f$ is the frequency of the noise. 
Though a full microscopic understanding of TLS noise is yet to be established, previous studies have shown that the TLS amplitude in MKIDs can be altered by changing the capacitor geometry \cite{noroozian2009two}, and through substrate surface treatments prior to the deposition of the metal films \cite{verjauw2021investigation}.
The generation-recombination noise of an MKID is caused by the continual breaking and reforming of Cooper pairs within the inductor. From the perspective of the resonator, the GR noise is also dependent on inductor geometry and should decrease with reduced inductor volume\cite{de2011number}. 
{The impact of amplifier noise can be minimized by using amplifiers with lower noise-equivalent temperatures or increasing the operating power and, therefore, the resulting signal-to-noise ratio. We typically operate at the power just below the bifurcation \cite{swenson2013operation} threshold, where the power-dependent nonlinear kinetic inductance starts to generate discontinuities in the resonance. }

In this paper, we fit detector noise power spectra to extract the TLS noise, the GR noise, and the amplifier white noise for a series of detector designs with different inductor volumes and interdigitated capacitor (IDC) geometries. 
We repeat this analysis with different detector materials, including niobium, aluminum, and manganese-doped aluminum (AlMn) with a lower superconducting transition temperature $T_c$ than aluminum. 
The goal of this exercise is to identify a set of geometric and material parameters that enable the tuning of detector noise to achieve the requirements of a given an experiment. This has immediate application both to our current efforts toward the development of SPT-3G+ \cite{dibert2022development} and to future missions with more stringent noise requirements such as an NEP at the level of $10^{-20}\mathrm{W\,/\sqrt{Hz}}$. 

\section{Device Design and Fabrication}
\label{sect:design}
\begin{figure}[!ht]
\centering
\includegraphics[width=0.95\linewidth]{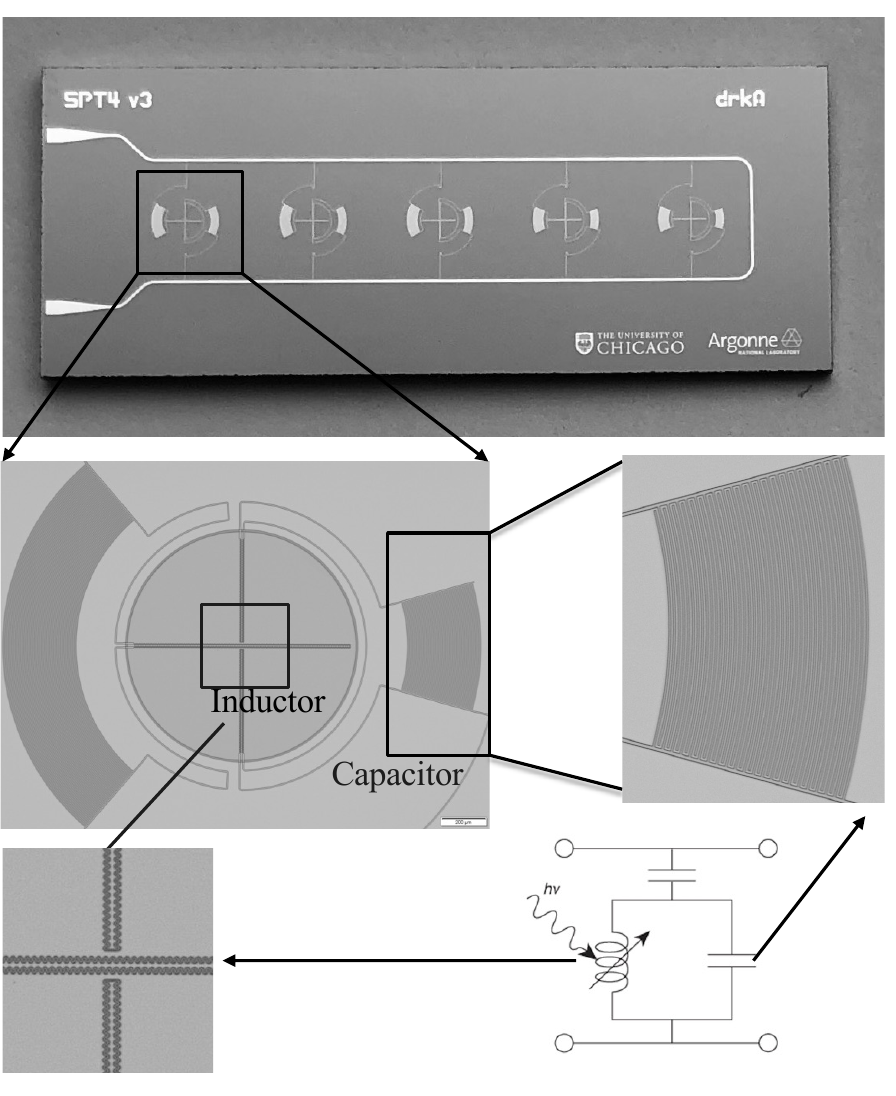}
\caption{Microscope images of the type (A) chip described in the text with different inductor volumes. Each pixel contains two detectors with two inductors oriented in orthogonal directions and two associated capacitors to form the resonators. The inductors are meandered metal lines, and the capacitors utilize the interdigitated design. The capacitors have curved tines that allow higher packing density in the final detector array. Bottom right diagram is from \cite{day2003broadband}.    }
\label{fig:design_structure}
\end{figure}

\begin{figure}[!ht]
\centering
\includegraphics[width=0.87\linewidth]{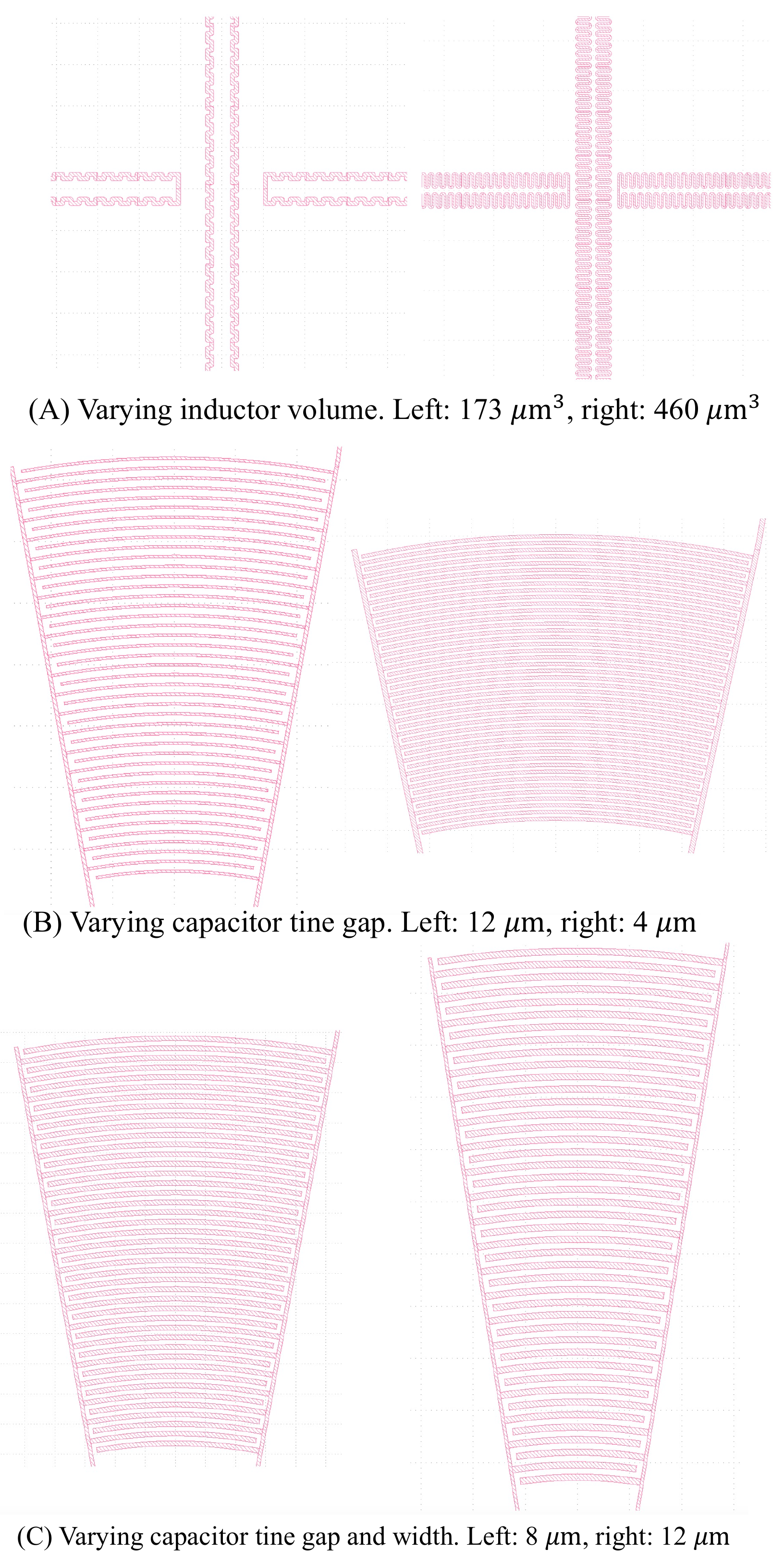}
\caption{Detector geometry variations. Here subfigures A), B), and C) correspond to designs A), B), and C) discussed in the text. }
\label{fig:design_geometries}
\end{figure}

The MKID design used for this study is based on the prototype {lumped element KID} design for SPT-3G+ \cite{dibert2022development}. We then modify component geometries and detector materials to explore their impact on detector noise. 
Figure~\ref{fig:design_structure} shows a photo of 1$\times$0.5~inch device used for this study.
The microstrip feedline is capacitively coupled to five pixels, each composed of two MKIDs aligned to orthogonal polarization optical modes.
Operating as detectors, these resonators are intended to couple to radiation via a feedhorn, which drives the design of the resonator inductor to double as an efficient photon absorber.
Each inductor is coupled to an IDC whose geometry sets the unique resonant frequency of each resonator on the device. 
Three types of devices were fabricated, each designed to test one change to device geometry.
These include: A) a device with varying inductor volumes for all five pixels to explore the impact on the GR noise, B) a device with a varying gap width between IDC tines or equivalently varying IDC filling factor, and C) a device with varying IDC tine width as well as the gap between tines. 
See Fig. \ref{fig:design_geometries} for the design geometries of A), B), and C). 
The devices are fabricated via a single-layer metal deposition on a high-resistivity silicon substrate. The resonator geometry is defined via optical photolithography and a wet etch. The silicon wafers are prepared with a buffered-HF etch to remove the native oxide layer prior to metal deposition.
Each six-inch wafer contains all of the geometric design variations, so that the film material remains as consistent as possible across all device types.

\section{Noise Measurement and Modeling}
\label{sect:modeling}
Detector noise timestreams were measured with a homodyne single-tone setup consisting of a microwave signal generator, quadrature demodulator, signal splitter, attenuators, and a 200~kHz commercial ADC.

{We measured our chips in a dilution refrigerator over a temperature range spanning $\sim 8-300$~mK. The input line has attenuators located at 4~K with a total attenuation of about 70~dB. The output signal is amplified by a cryogenic low-noise amplifier with $\sim$ 30~dB gain and 6~K noise temperature, as well as a room-temperature amplifier with $\sim$30~dB gain. }

\begin{figure}[!ht]
\centering
\includegraphics[width=0.95\linewidth]{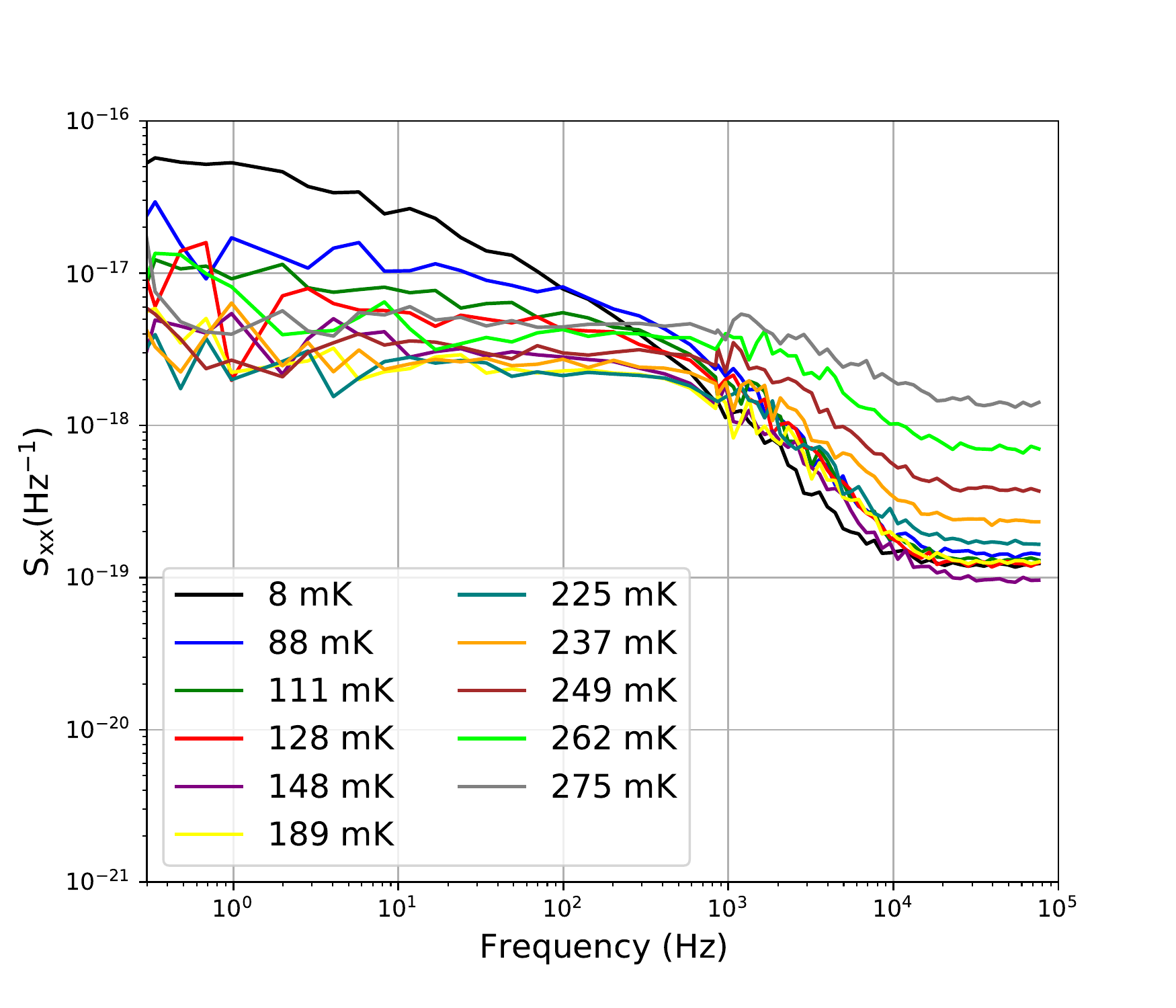}
\caption{Sample fractional frequency shift noise PSD for a resonator. The $1/\sqrt{f}$ feature at lower stage temperature corresponds to TLS noise and decreases as we increase the operating temperature. The flat white noise level in the mid-frequency range corresponds to the generation and recombination of quasiparticles. The white noise floor above the GR rolloff at a few kHz is mostly system noise dominated by amplifier noise.  }
\label{fig:sample_noise_psd}
\end{figure}

 We report the noise power spectral densities (PSDs) calculated using the local-gradient procedure in \cite{barry2014development} and we have verified the consistency of this method with the phase-shift method described in the same reference. 
 Figure \ref{fig:sample_noise_psd} shows example noise measurements for one detector taken at a microwave power on the feedline of -110 dBm.
 At 8~mK, we observe in the noise PSD a clear $1/\sqrt{f}$ slope associated with TLS noise \cite{zmuidzinas2012superconducting} that decays with increasing device temperature. Around 1000~Hz, we observe a rolloff in the PSD that is related to the quasiparticle lifetime. The frequency of this rolloff shifts to a higher frequency as the operating temperature increases {since the quasiparticle lifetime reduces at higher temperatures (Eq. \ref{eq:tau_dependence})} \cite{de2012generation}.
 The noise beyond ~20~KHz is dominated by the white noise level of the cryogenic amplifier. {At higher temperatures, the detector's $dI/df$ and $dQ/df$ responsivities reduce, which causes the white noise level in fractional frequency shift ($df/f$) to increase when we convert the demodulated time-ordered data to the $df/f$. We allow parameter $C$ to vary when fitting data at different temperatures.}  

Each noise PSD is fit with the following model \cite{mcgeehan2018low}:
\begin{equation}
S_{x x}(f)=\left(\frac{A+B f^{-n}}{1+(2 \pi f \tau)^2}+C\right),
\label{fit_model}
\end{equation}
where $S_{x x}$ is the frequency noise PSD, $A$ is the GR noise component, $Bf^{-n}$ is the $1/\sqrt{f}$ noise (mostly dominated by TLS in our system), $\tau$ is the quasiparticle lifetime, and $C$ is the system white noise level. 

To break the degeneracy of parameters, we note that the GR noise level $A$ does not depend on temperature {because the temperature dependence of $N_{qp}$ and $\tau$ cancels \cite{de2011number}}.
Thus, we can fit for $A$ using high-temperature data where the TLS component is suppressed, and then fit for $B$ in low-temperature data using a model with fixed $A$.
We found that our low-temperature data is consistent with $n\sim 0.5$ and therefore fixed $n=0.5$ to better expose the geometry dependence of parameter $B$.
The quasiparticle lifetime, $\tau$, and the system white noise level, $C$, are not degenerate with other parameters since they correspond to distinct features: the rolloff and the flat region at high frequency ($\sim10^5$~Hz).

Some of the AlMn detectors were found to have a non-negligible resonator ring-down time $\tau_\text{res}={Q}/{\pi f_0}$, where $Q$ is the resonator quality factor, and $f_0$ is the resonant frequency.
Since the ring-down time is assumed to be negligible in Eq. (\ref{fit_model}), we re-introduced the rolloff term $1+(2\pi f\tau_\text{res})^2$ to the denominator of Eq. (\ref{fit_model}) for AlMn detectors with $\tau_{res}$ from independent measure of $Q$ and $f_0$ \cite{de2011number}.
The AlMn resonators also have low bifurcation power \cite{lisovenko2022characterization}, which in some cases was only slightly above the system white noise floor. 
As a result, we collected limited data for a fraction of AlMn resonators within a small range of stage temperature and bias powers relative to the datasets taken for other materials. 
For a robust extraction of the fit parameters, we fit multiple noise curves taken for the same resonator at different temperatures and powers simultaneously.

\section{Generation-recombination Noise Optimization}

The generation-recombination noise contribution to the quasiparticle fluctuation PSD is expected to follow:

\begin{equation}
S_N(f) = \frac{4N_{qp}\tau}{1+(2\pi f \tau)^2}, 
\end{equation}
where $N_{qp}$ is the number of quasiparticles. The numerator of this expression should scale with detector volume $V_L$ since $N_{q p}=n_{q p}V_L$, where $n_{qp}$ is the quasiparticle number density. We seek to demonstrate the dependence of the numerator on inductor volume.

\begin{figure}[!ht]
    \centering
  \subfloat[ X-polarization detectors, aluminum, type A \label{volume_a}]{%
       \includegraphics[width=0.8\linewidth]{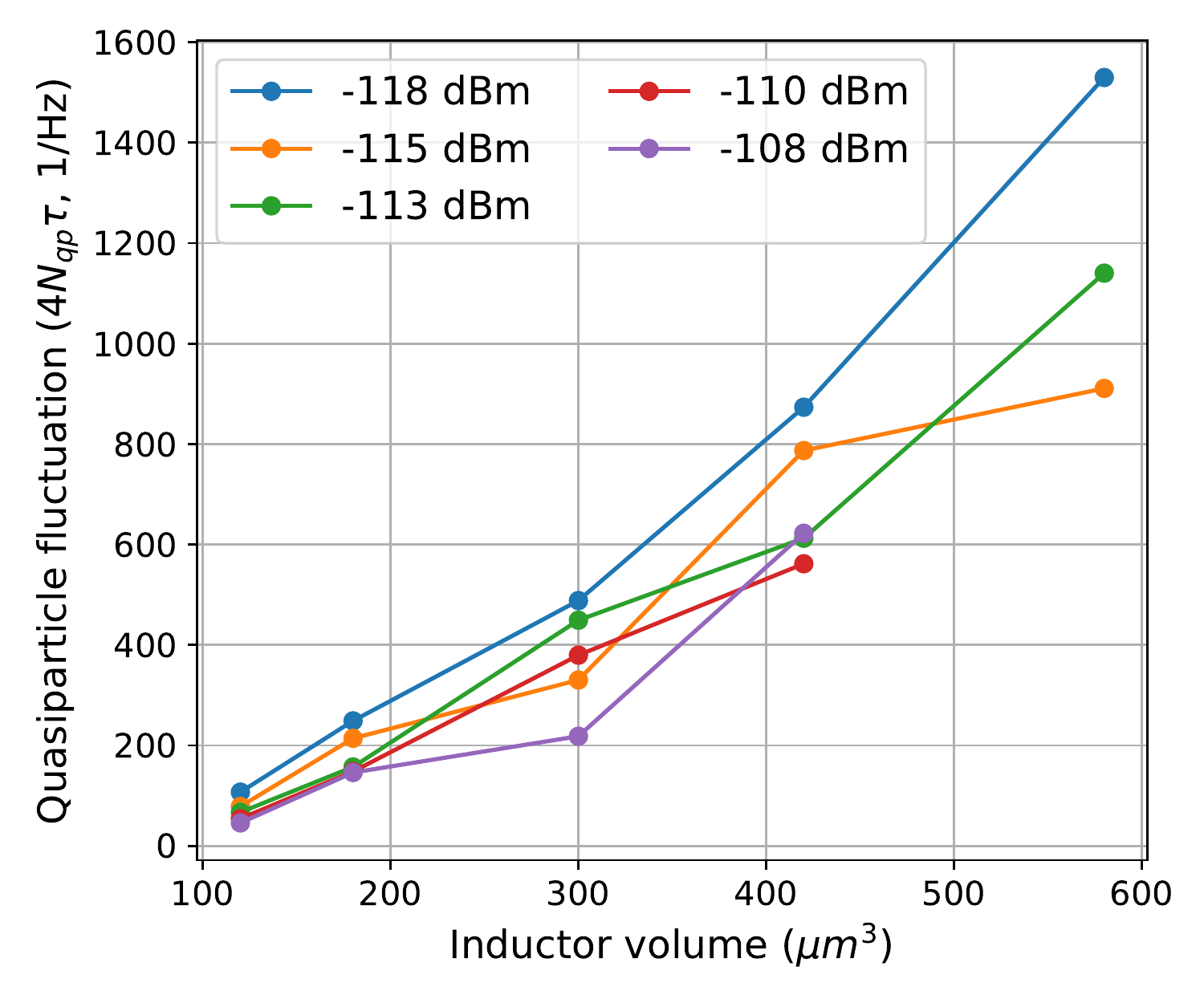}}
    \hfill
  \subfloat[ Y-polarization detectors, aluminum, type A\label{volume_b}]{%
        \includegraphics[width=0.8\linewidth]{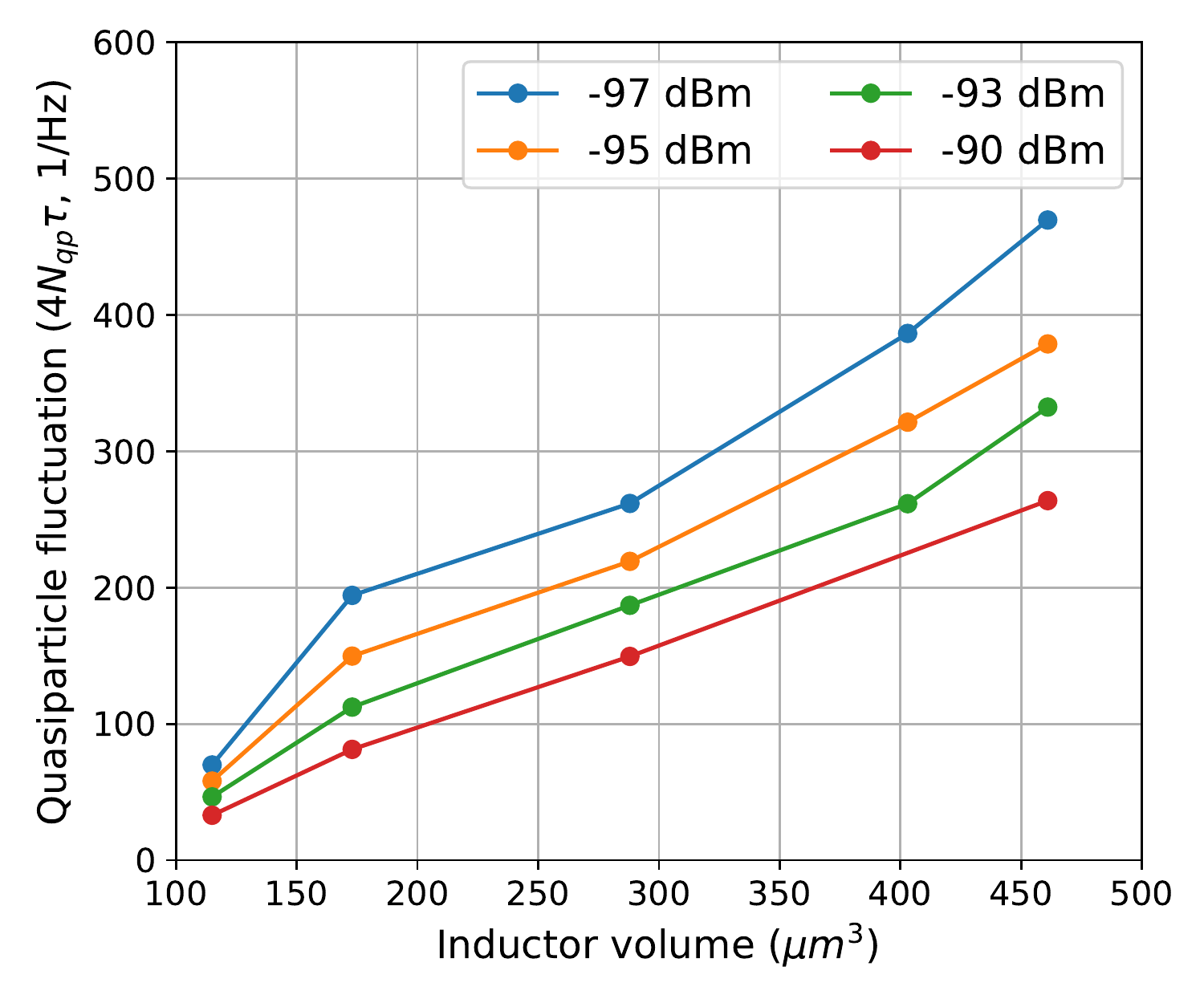}}
    \\
  \caption{Quasiparticle number fluctuation PSD vs. inductor volume for X and Y polarizations. We separate X and Y due to different inductor geometries.}
  \label{fig:ind_volume} 
\end{figure}

We fit the noise PSDs of aluminum detectors with varying inductor volume (Type A from Section \ref{sect:design}), and extracted the GR noise level ($A$ in Eq. (\ref{fit_model})). This parameter theoretically has dependence $A=4 N_{q p} \tau\left(d x / d N_{q p}\right)^2$, where $x=df/f$. To isolate the linear dependence of $4 N_{qp} \tau$ on inductor volume, we divide out the $d x / d N_{q p}$ term by recognising that
\begin{equation}
    \frac{dx}{dN_{qp}} = \frac{dx}{dT} \frac{dT}{dN_{qp}} = \frac1 {V_L} \frac{dx}{dT} \frac{dT}{dn_{qp}},
\end{equation}
and making use of expression for quasiparticle number density:
\begin{equation}
\label{eq:nqp}
n_{q p}=2 N_0 \sqrt{2 \pi k_B T \Delta} \exp \left(-\Delta / k_B T\right),
\end{equation}
where $N_0$ is the single spin density of states at the Fermi level ($1.72 \times 10^{10} \mathrm{\mu m}^{-3} eV^{-1}$ for aluminum), $k_B$ is the Boltzmann constant, and $\Delta = 1.76k_BT_c$ is the energy gap for aluminum.
Fig. \ref{fig:ind_volume} shows $4 N_{qp} \tau$ as a function of inductor volume for the aluminum devices. {The data was taken at 0.2~K where the TLS noise is reduced}. We grouped the detectors by polarization alignment, as the inductor geometry differs slightly between polarization. This figure shows that the GR noise has a positive correlation with inductor volume. We note that X and Y detectors are different (Fig. \ref{fig:design_geometries}A). 
{The observed differences between X vs. Y polarizations are consistent with different excess quasiparticles arising from different coupling to the RF bias between the two classes of detectors. The coupling quality factors $Q_c$ for X detectors are $(1.0\pm0.1)\mathrm{e}5$ and $(1.1\pm0.4)\mathrm{e}4$ for Y detectors, which are different by a factor of $\sim$10. The difference in $Q_c$ leads to a difference in bifurcation power of $\sim$20~dB. With a similar amount of power at the feedline, X detector noise may have more contribution from quasiparticles agitated by the bias power that makes the noise for X detectors deviate from that for Y and a linear dependence on volume.  As to the power dependence for Y, we noted that Y detectors have more TLS noise contribution compared to X detectors, probably due to different internal power for Y or the different geometry. And the sloped TLS noise can have some degeneracy with the flat GR noise term. As we increased the power, we saturate more of the TLS component and reduce the systematics from the TLS component, leading to a reduction of fit A values. }


\section{Two-level System Noise Optimization}

We fit noise PSDs for aluminum detectors of varying IDC tine gap width (Type B from \ref{sect:design}) and IDC tine width (Type C from \ref{sect:design}) to extract the TLS amplitude ($B$ from Equation \ref{fit_model}).

\begin{figure}[!ht]
    \centering
  \subfloat[ Fit $B$ vs. capacitor finger gap, aluminum, type B  \label{vs_gap_width}]{%
       \includegraphics[width=0.78\linewidth]{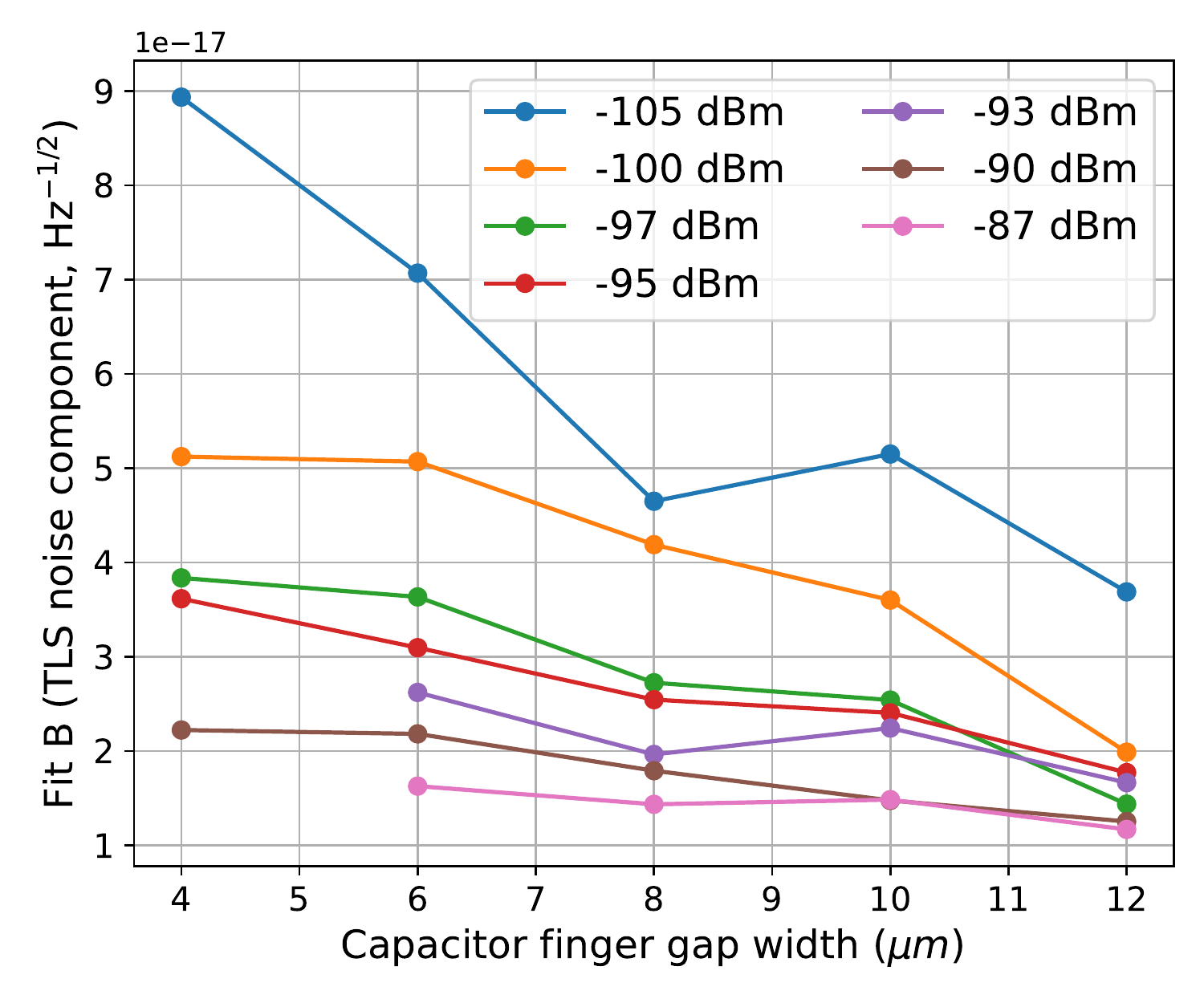}}
    \hfill
  \subfloat[ Fit $B$ vs. $F\delta$, aluminum, type B \label{vs_fd}]{%
        \includegraphics[width=0.8\linewidth]{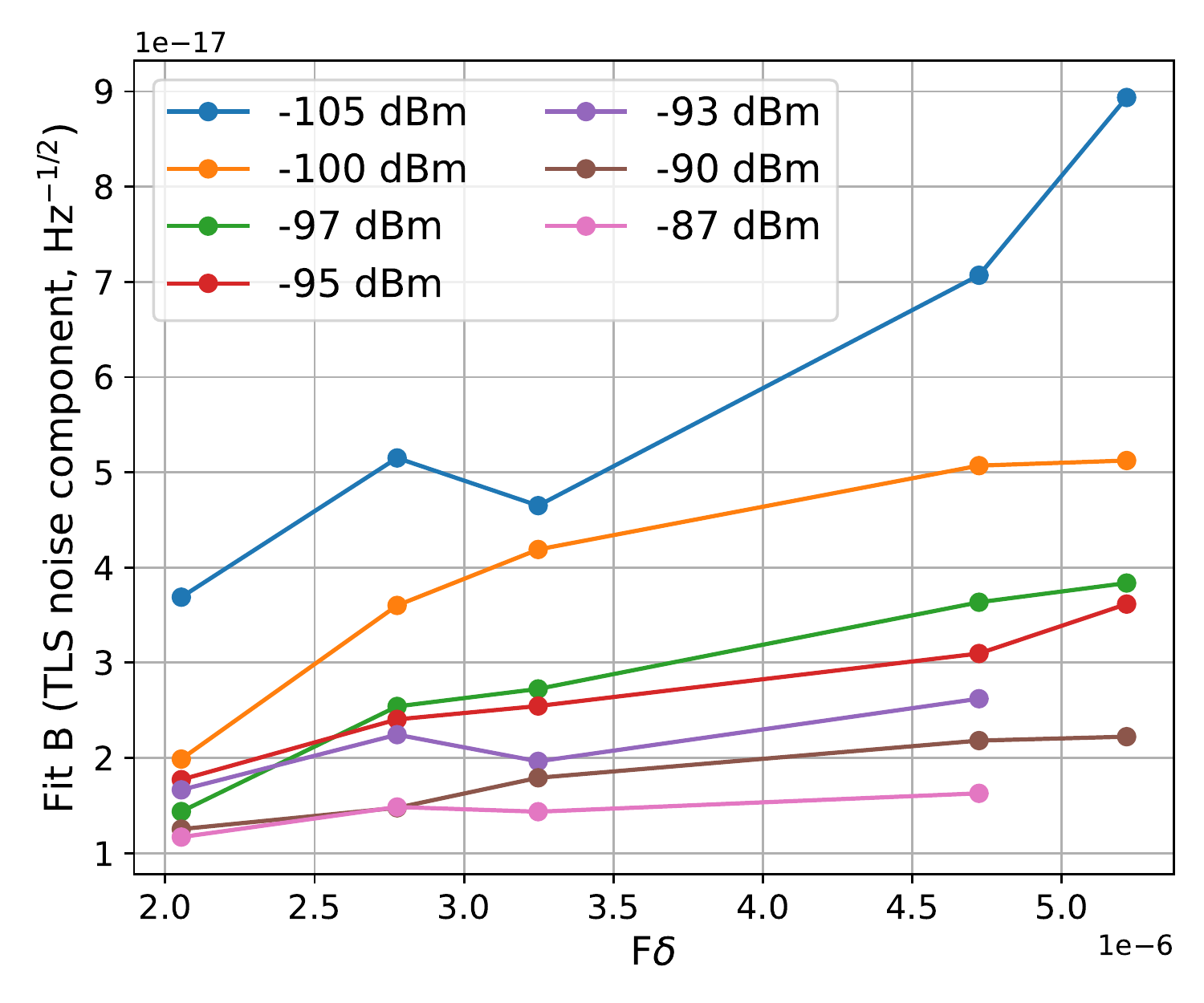}}
    \hfill
  \subfloat[ Fit $B$ vs. $F\delta$, niobium, type B \label{vs_fd_nb}]{%
        \includegraphics[width=0.8\linewidth]{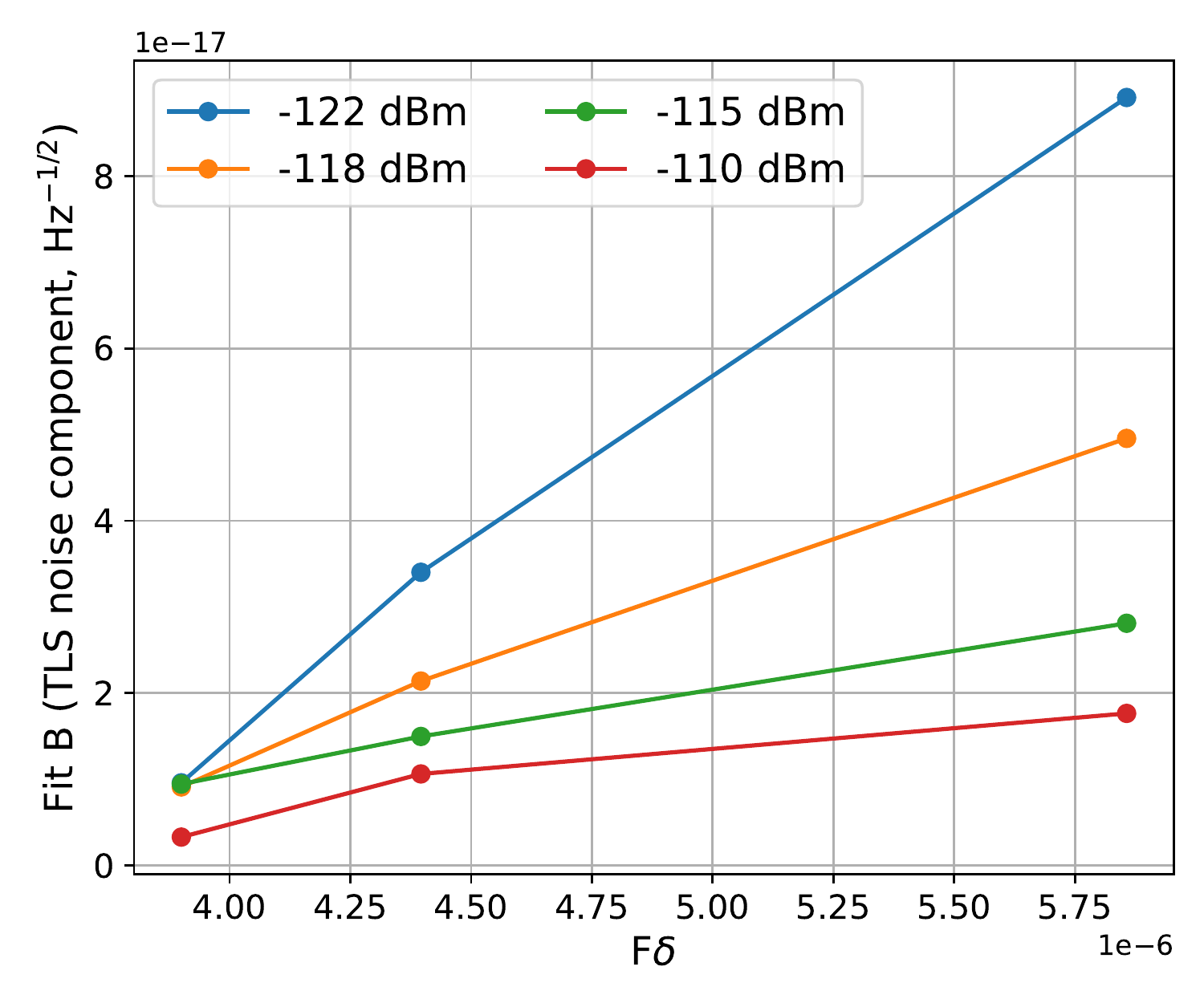}}
  \caption{TLS noise's dependence on capacitor finger gap width and $F\delta$. (a) and (b) are for aluminum, and (c) is for niobium. }
  \label{fig:tls_noise_dependence} 
\end{figure}

Fig.~\ref{vs_gap_width} shows a clear decrease in TLS amplitude as the IDC tines are spaced further apart. With a larger gap between the IDC tines, the electric field is smaller, resulting in weaker coupling to the dipole states in the amorphous solid.

Coupling of the field to the TLS is expected to cause a shift in both quadratures of the dielectric response function \cite{pappas2011two}. {Assuming a log-uniform distribution of two-level tunneling states, the resonance frequency shift can be calculated from the modification to the dielectric \cite{gao2008equivalence}.}
We can measure the quantity $F\delta$ by fitting resonant frequency vs. temperature to a TLS model \cite{zmuidzinas2012superconducting}. 
Here $F$ is the filling factor, which defines the fraction of the power coupled through the dielectric, and $\delta$ is the loss {tangent} for the dielectrics. 
We measured the TLS loss for the resonators and plotted the fit $B$ values vs. $F\delta$ in Fig. \ref{vs_fd}, where a clear correlation is observed. 
We repeated these measurements for niobium resonators with detector type C and found a similar positive correlation between $B$ and $F\delta$ in Fig. \ref{vs_fd_nb}. 

To minimize TLS noise, we should reduce the filling factor in dielectrics or surface states. While intuitive ways such as changing capacitor gap size exist, a more systematic way for this optimization is through simulations {using electromagnetic simulation software, such as HFSS \footnote{https://www.ansys.com/products/electronics/ansys-hfss}}. We can simulate the filling factors for potential TLS locations such as dielectrics, surface oxide layers, or interface layers and study their correlation with the observed TLS noise levels. 

\begin{figure}[!ht]
    \centering
  \subfloat[ Fit $B$ vs. driving power, aluminum, type B  \label{vs_power}]{%
       \includegraphics[width=0.9\linewidth]{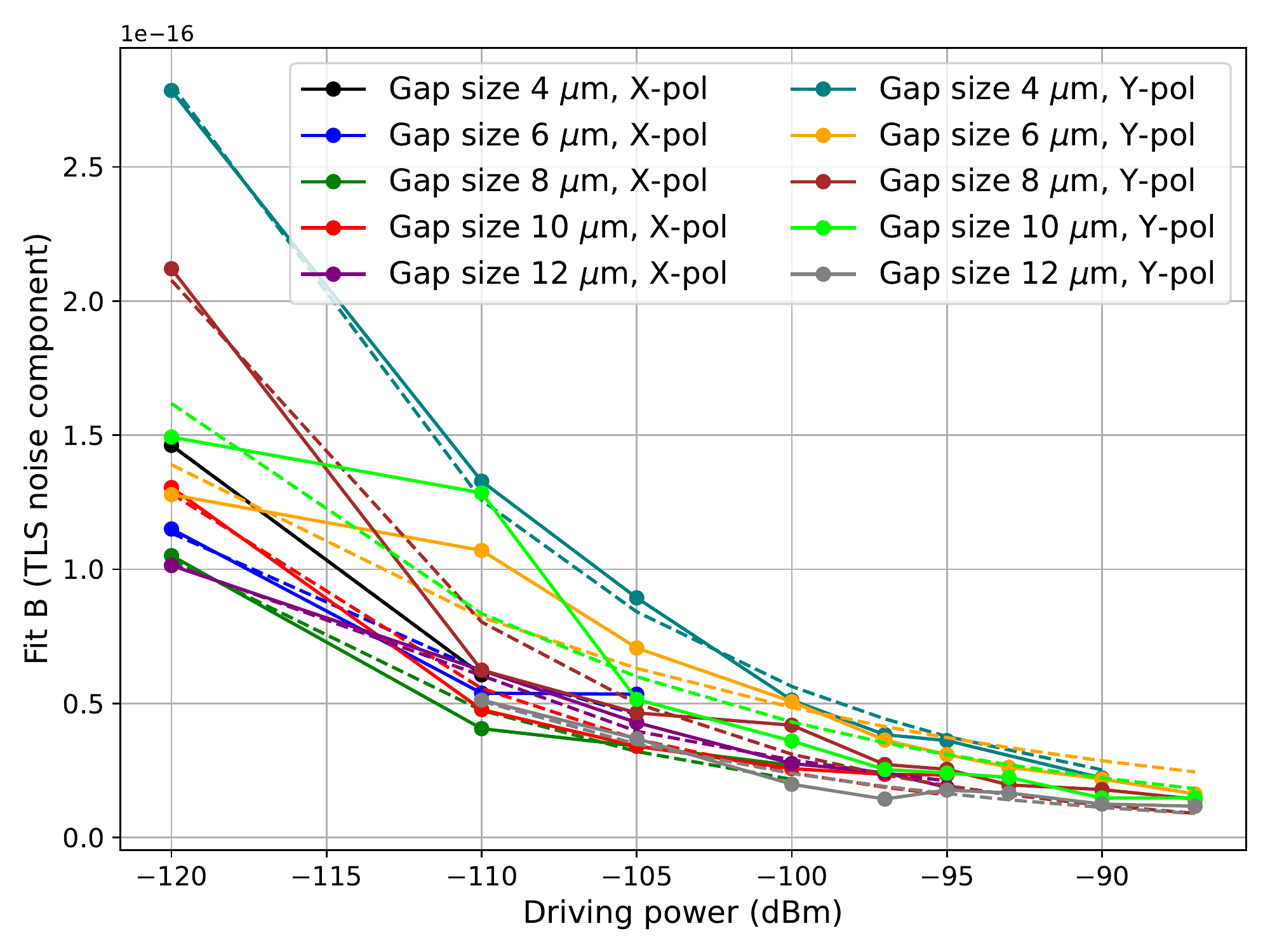}}
    \hfill
  \subfloat[ Fit $B$ vs. operating temperature, aluminum, type B \label{vs_temperature}]{%
        \includegraphics[width=0.9\linewidth]{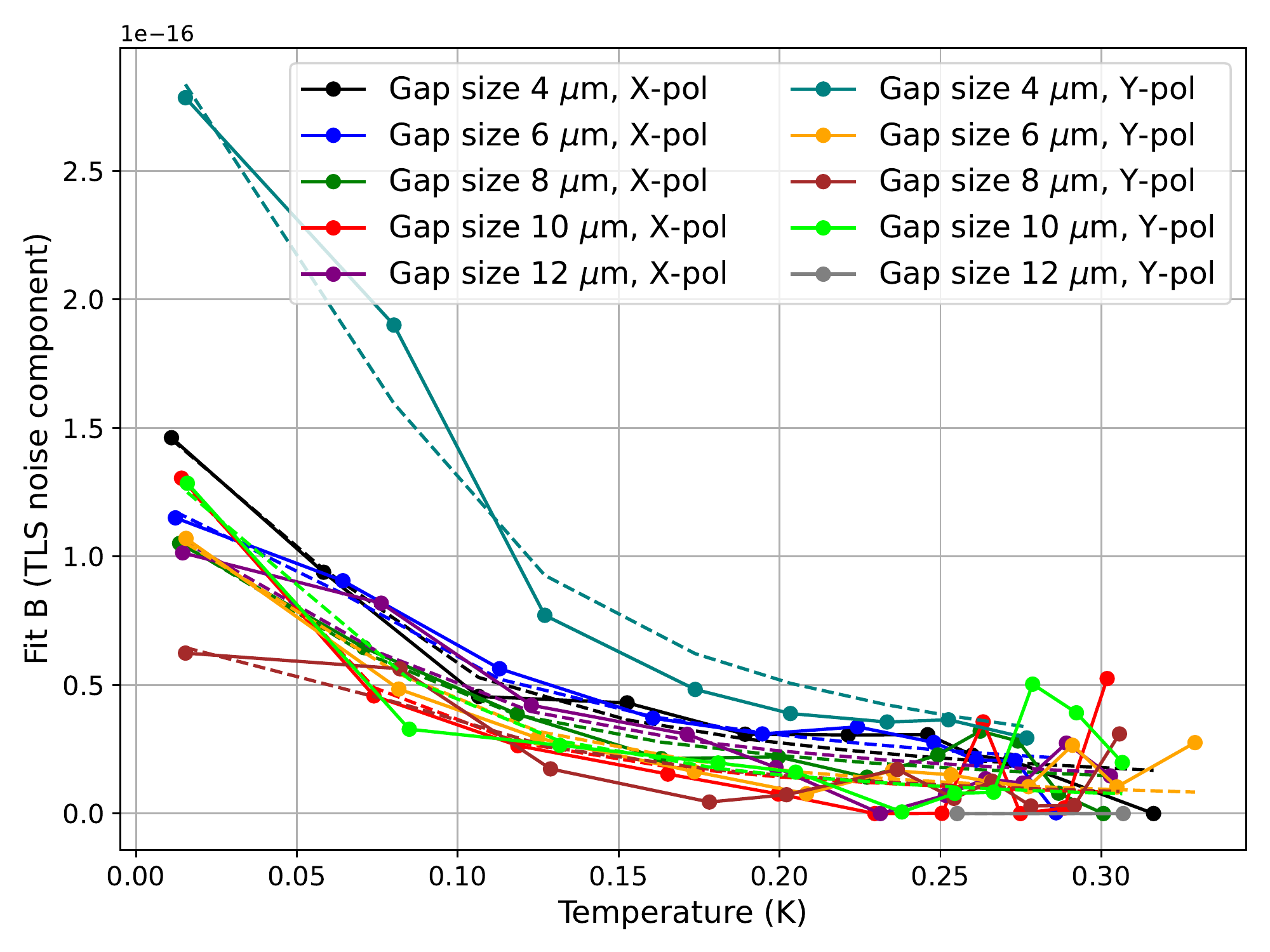}}
  \caption{TLS noise's dependence on driving power and temperature. We have included ten detectors in each plot for five different gap sizes and two polarizations. The design is type (B) in Fig. \ref{fig:design_geometries}. We added dashed fit lines to Eq. \ref{eq:noise_dependence} and got $\alpha=-0.3\pm 0.1$ and $\beta=0.8\pm 0.2$.  }
  \label{fig:tls_noise_dependence_power_temperature} 
\end{figure}

There are additional external parameters for tuning the TLS noise, such as the driving power and the operating temperature. The phenomenological TLS model predicts the dependence on power and temperature as \cite{kumar2008temperature}
\begin{equation}
\label{eq:noise_dependence}
S_{xx} \propto P^{\alpha} f^{-1/2} T^\beta \tanh \left(h f_0 / 2 k_B T\right),
\end{equation}
where $P$ is the power, $f$ is the frequency, $\beta$ is an empirically derived exponent, and $f_0$ is the resonant frequency. We fit our measured TLS noise levels parametrized by $B$ {qualitatively} to the model in Eq. (\ref{eq:noise_dependence}) and plot the data and dashed fit lines in Fig. \ref{fig:tls_noise_dependence_power_temperature}. {The fit values are $\alpha=-0.3\pm 0.1$ and $\beta=0.8\pm 0.2$. Similar to \cite{kumar2008temperature}, our data cannot distinguish temperature dependence between this model and a power law of $T^{-0.6}$. Ref. \cite{kumar2008temperature} reported $\alpha=-0.46\pm0.05$ and $\beta=-0.14\pm0.02$ for niobium and the temperature range of 0.12 to 1.2~K. Our measurements for aluminum from 0.01 to 0.32~K indicate the exponents can change with material and temperature.  We note that the physical mechanism of the TLS noise remains unclear and further study is needed beyond the phenomenological model. }To reduce TLS noise, we can operate at a high bias power just below the onset of bifurcation \cite{swenson2013operation}, and raise the stage temperature ensuring that the white noise level and resonator quality factor are not degraded.

\section{Quasiparticle lifetime}

\begin{figure}[!ht]
\centering
\includegraphics[width=0.95\linewidth]{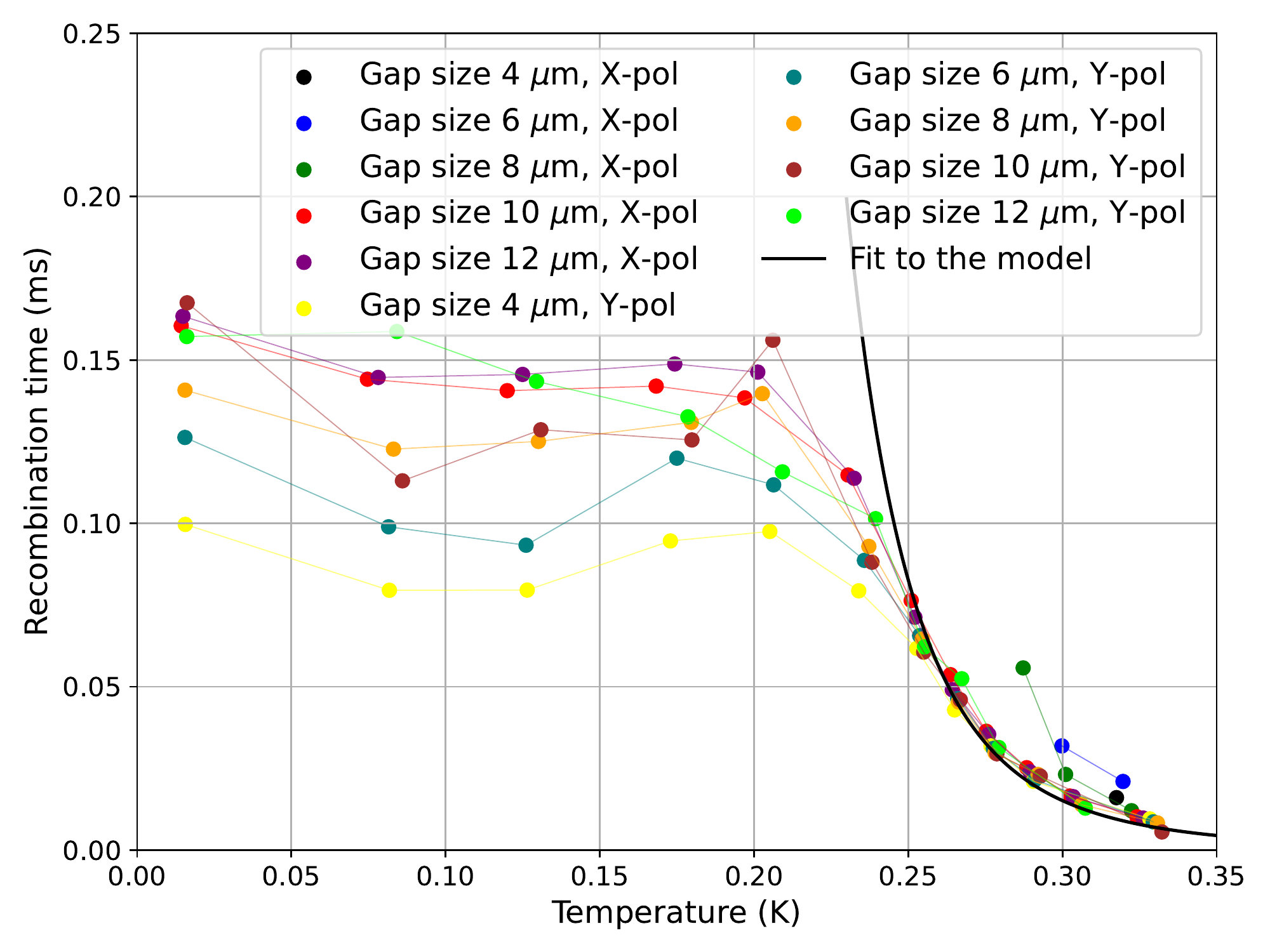}
\caption{Quasiparticle lifetime vs. operating temperature. We have included ten detectors in each plot for five different gap sizes and two polarizations. Plotted is for aluminum type (B) in Fig. \ref{fig:design_geometries}. We fit data above 0.24~K to the model in Eq. \ref{eq:tau_dependence} with the correction. }
\label{fig:time_constant_vs_T}
\end{figure}

The quasiparticle lifetime is another parameter we can fit from the $S_{xx}$ rolloff frequency. {It is related to the recombination time $\tau_R$, which for low temperatures is approximately given by \cite{kaplan1976quasiparticle}.}  The recombination time is summarized in \cite{kaplan1976quasiparticle}:  

\begin{equation}
\label{eq:tau_dependence}
\tau_R=\frac{\tau_0}{\sqrt{\pi}}\left(\frac{k_B T_c}{2 \Delta}\right)^{5 / 2} \sqrt{\frac{T_c}{T}} e^{\Delta / k_B T}=\frac{\tau_0}{n_{q p}} \frac{N_0\left(k_B T_c\right)^3}{2 \Delta^2}, 
\end{equation}
where $\tau_0$ is a material-dependent characteristic electron-phonon interaction time {and can be modified by impurity scattering}. Note that one recombination is associated with the disappearance of two quasiparticles so that $\Gamma_R^*=2\Gamma_R F_{\omega}^{-1}$\cite{wilson2004quasiparticle}, where $\Gamma_R$ is the recombination rate, $\Gamma_R^*$ is the time constant for a small quasiparticle perturbation to decay (quantity measured experimentally), and $F_{\omega}$ is the phonon trapping factor that accounts for pair breaking by emitted phonon from the recombination. If we neglect phonon trapping, $\Gamma_R^*=2\Gamma_R$, and the resulting measured quasiparticle lifetime $\tau=\tau_R/2$, where $\tau_R$ is defined in Eq. (\ref{eq:tau_dependence}). We plot the dependence of the quasiparticle lifetime as a function of temperature in Fig. \ref{fig:time_constant_vs_T} and then fit it to the model in Eq. (\ref{eq:tau_dependence}) with the factor of two correction using temperatures above 0.24~K. The fit value for $\tau_0$ is 0.3$\pm$0.1~$\mu$s, which is of the same order of magnitude as for previous measurements \cite{kaplan1976quasiparticle, wilson2001time} but somewhat smaller. Ref. \cite{barends2009enhancement} found impurities can change quasiparticle recombination time, which could mean there are impurities in our aluminum. Another uncertainty comes from $T_c$ and $\Delta$. We found the $\tau_0$ in the model fit sensitive to the preset $T_c$ and $\Delta$ values, and that using $T_c=1.2$~K and $\Delta = 1.76k_B T_c$ gives $\tau_0=0.6\,\mu$s for the fit. The current $T_c=1.33$~K we used for this analysis is from a four-wire temperature sweep, which is close to our measurement of 1.37~K by fitting $f$ vs. $T$ to a Mattis-Bardeen model \cite{gao2008equivalence}. In the model, $\Delta = 1.76\,k_B T_c$ may not be accurate for a thin film, and we neglected phonon trapping. We note these uncertainties for our $\tau_0$ result. The quasiparticle lifetime below $\sim$0.2~K is limited by other relaxation mechanisms, such as nonequilibrium quasiparticle excitations \cite{martinis2009energy, catelani2011quasiparticle}. 

\section{AlMn Noise Results}

The AlMn samples we used are the same as the ones reported in \cite{lisovenko2022characterization}.
We can reach a lower $T_c$ than pure aluminum by doping aluminum with manganese, and baking provides yet another knob to tune their $T_c$.
Low-Tc MKIDs are interesting for spectrometers because they open up below 100~GHz for a few redshifted CO lines and are also useful for MKIDs-based dark matter searches that require higher sensitivity.
	The aluminum and AlMn films were sputtered using high-purity aluminum and Mn doped aluminum with 1050~ppm (parts per million) and 1150~ppm  doping levels, respectively.
The superconducting transition temperatures for the aluminum sample, 1050~ppm AlMn, and 1150~ppm AlMn are 1.37~K, 0.73~K, and 0.61~K, respectively.
The 1050 ppm sample subsequently went through a baking process at 180~$^{\circ}\mathrm{C}$ for 10 minutes, which shifted its $T_c$ from 0.73~K pre-baking to 0.78~K after baking.
Our goals with the AlMn samples are to understand the dependence of GR noise level and quasiparticle lifetime on the doping and baking conditions.

The measured noise PSDs for the AlMn resonators are similar to those for the aluminum resonators shown in Fig.~\ref{fig:sample_noise_psd}, but the PSD curves are noiser especially in the low frequency region below about 1~kHz. 
	Since the GR noise level $A$, in the temperature and bias power range where we took data,  is independent of temperature or bias power, we fit the set of PSD curves for an AlMn resonator simultaneously with a single parameter $A$, while the $\tau_{res}$ corrections were applied independently to each noise curve.
	The quasiparticle lifetime for each noise curve, $\tau$, was also obtained from this global fit.
	The GR noise level can in principle be expressed in quasiparticle number fluctuations, $4 N_{qp} \tau$, by factoring out $dx/dN_{qp}$ from $A$.
	However, we did not do this for the AlMn resonators because $N_0$ for AlMn was not known to us.

\begin{figure}[!ht]
\centering
\includegraphics[width=0.95\linewidth]{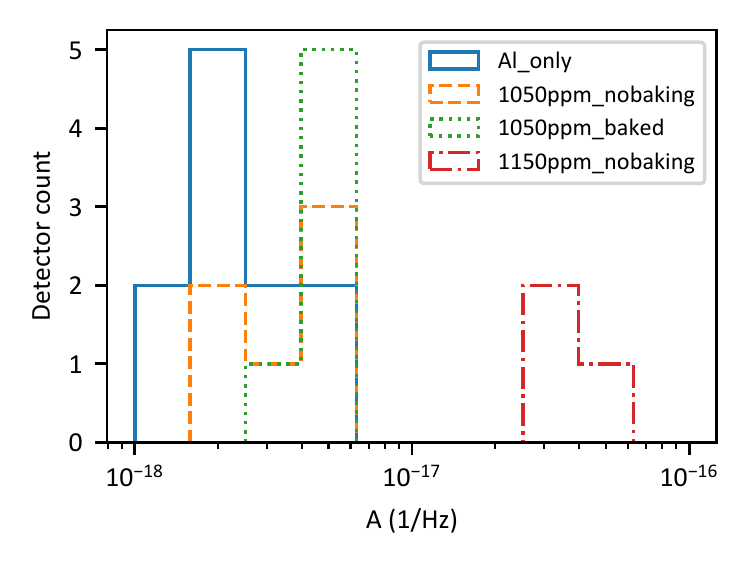}
\caption{Histograms of the GR noise level $A$ in $S_{xx}$ for pure aluminum and AlMn with different Mn concentrations. }
\label{fig:A_hist}
\end{figure}

Fig. \ref{fig:A_hist} shows the GR noise level $A$ in fractional frequency shift for different doping and baking configurations.
	We observe an increase in GR noise with increasing Mn doping levels, and we find that baking makes the GR noise slightly worse for the 1050~ppm AlMn resonators.
	Clearly, these observations are still limited by statistics because of the limited dataset.

\begin{figure}[!ht]
\centering
\includegraphics[width=0.95\linewidth]{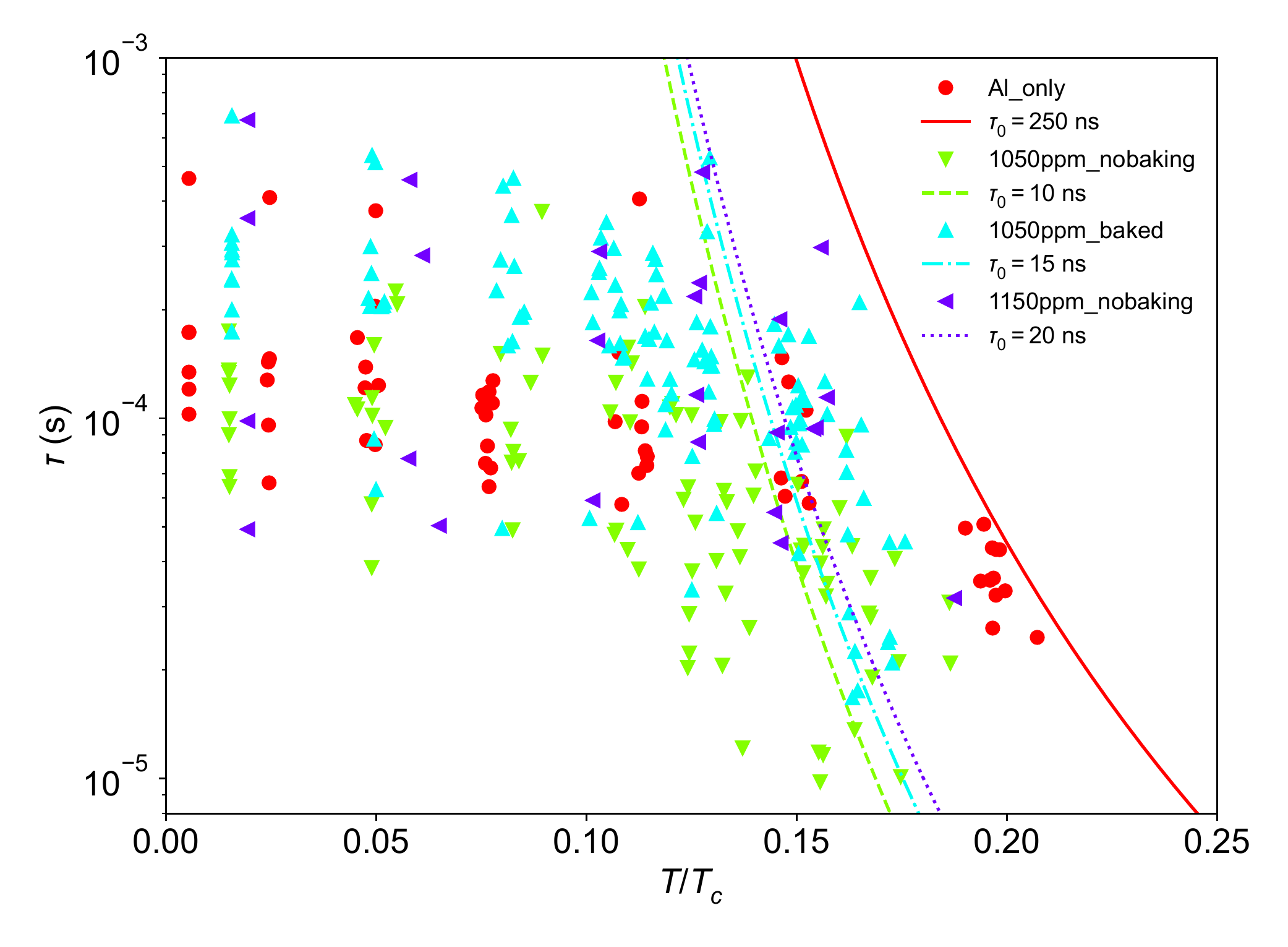}
\caption{Quasiparticle lifetime vs. reduced temperature $T/T_c$ for pure aluminum and AlMn with different Mn concentrations.
    We have added model curves using  Eq. (\ref{eq:tau_dependence}) with the samples' measured $T_c$.
    The $\tau_0$ values for the model are chosen such that the curves follow the falling edges at higher temperatures.
    The data is too noisy for a good fit, but we can still tell $\tau_0$ for AlMn is an order of magnitude smaller than pure aluminum if we assume a BCS description.  }
\label{fig:time_constant_compiled}
\end{figure}

We plot the fitted quasiparticle lifetime $\tau$ as a function of the reduced temperature $T/T_c$ in Fig. \ref{fig:time_constant_compiled}.
	All samples show saturated $\tau$ when $T/T_c<0.1$, but their rolloff start at different reduced temperatures.
	Doped aluminum with impurities can be analyzed with more complex theories developed by Zittartz, Bringer, and Müller-Hartmann \cite{zittartz1972impurity}, or Kaiser \cite{kaiser1970effect}.
	Barends et al. \cite{barends2009enhancement} obtained some qualitative conclusions applying these models but did not model their data quantitatively.
	$\tau$ for AlMn in Fig. \ref{fig:time_constant_compiled} show wider scatter compared with the Al resonators shown in Fig.~\ref{fig:time_constant_vs_T}.
	This is because  the AlMn resonators have lower bifurcation power than pure aluminum, which required a reduction in the bias power for the measurement and consequently {decrease the signal-to-noise level}.

	We can draw qualitative conclusions from the $\tau$ vs. $T/T_c$ dependence shown in Fig. \ref{fig:time_constant_compiled}.
	In the same figure, the modeled quasiparticle lifetime $\tau=\tau_R/2$ using Eq. (\ref{eq:tau_dependence}) are also shown.
	$\tau_0$'s are chosen that the theoretical curves match the corresponding roll-off edges visually.
	Assuming all samples are BCS superconductors, the samples doped with manganese have $\tau_0$ about an order of magnitude smaller than pure aluminum.
	This reduction in  $\tau_0$ is not unexpected through its dependence on impurities~\cite{rammer1986destruction}.
	We realize that the BCS modeling may not fully apply here: O'Neil et al.~\cite{o2010quasiparticle} demonstrated that the density of states of AlMn remains BCS-like, but additional sub-gap states and gap edge smearing are also present.

\begin{figure}[!ht]
\centering
\includegraphics[width=0.95\linewidth]{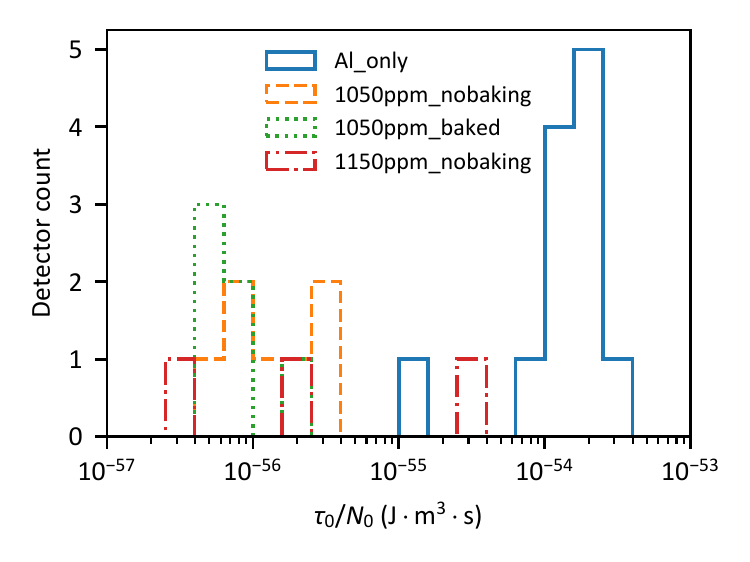}
\caption{$\tau_0/N_0$ histograms for pure aluminum and AlMn with different Mn concentrations.}
\label{fig:tau0_N0_hist}
\end{figure}

Finally, we use the GR noise level $A$ to constrain $\tau_0$ and $N_0$.
With Eq. (\ref{eq:nqp}) and Eq. (\ref{eq:tau_dependence}), we can decompose the dependence of $A$ into:
\begin{equation}
\label{eq:a_dependence}
 A = 4 N_{q p} \tau\left(d x / d N_{q p}\right)^2 = \frac{\tau_0}{N_0} \frac{2}{V_L} \left(\frac{dx}{d\rho}\right)^2  \frac{(k_BT_c)^3}{2\Delta^2}, 
\end{equation}
 where $\rho=n_{qp}/N_0 = 2\sqrt{2\pi k_B\Delta}\exp{(-\Delta/k_BT)}$.
 We can put constraints on $\tau_0/N_0$ using Eq. (\ref{eq:a_dependence}) since all other terms can be measured or are known from the design.
 The resulting histograms for $\tau_0/N_0$ are summarized in Fig. \ref{fig:tau0_N0_hist}. 
 The $\tau_0/N_0$ values for pure aluminum are about two orders of magnitude larger than those for AlMn, indicating that the density of states $N_0$ for AlMn is higher than that for aluminum since the $\tau_0$'s only differ by one order of magnitude. Ref \cite{deutz1981local} reported a calculation of local density of states for Al with impurities. 
 One caveat is that we have assumed all the samples are BCS-like.
 
\section{Conclusions}
In this paper, we explored and validated a few directions to control the noise components of MKIDs. For aluminum detectors, we were able to tune the GR noise by changing the inductor volume of our detectors and observed a positive correlation between the quasiparticle fluctuation noise and inductor volume. We can reduce the TLS noise using three methods: 1) lowering the capacitor filling factor, 2) raising the operating temperature, and 3) increasing the driving power. We compared the temperature and power dependence of TLS with models and correlated the TLS noise with $F\delta$ measured using a TLS dielectric loss model. We did the TLS study for both niobium and aluminum resonators. We also explored the noise components' dependence on manganese doping with AlMn detectors. We found manganese doping increases the GR noise level. Assuming a BCS description, we observed that AlMn samples prefer a lower $\tau_0$ and a higher $N_0$ than aluminum samples. The noise for our samples can be controlled below the photon noise limit for ground-based photometer applications (e.g., SPT3G+). However, we need more thorough optimization for space applications or spectrometers with narrower bands that require lower noise. This study provides a few directions for future noise optimization.

\section{Acknowledgements}
Work at Argonne, including the use of the Center for Nanoscale Materials, an Office of Science user facility, was supported by the U.S. Department of Energy, Office of Science, Office of Basic Energy Sciences, and Office of High Energy Physics, under Contract No. DE-AC02-06CH11357. Zhaodi Pan is supported by ANL under the award LDRD-2021-0186. Maclean Rouble acknowledges funding from the Natural Sciences and Engineering Research Council of Canada and Canadian Institute for Advanced Research.

\bibliographystyle{IEEEtran}
\bibliography{paper_reference.bib}



\end{document}